# Model Based Software Development: Issues & Challenges

N Md Jubair Basha[1], Salman Abdul Moiz[2] & Mohammed Rizwanullah[3]

[1&3]IT Department, Muffakham Jah College of Engineering & Technology, Hyderabad, India
[2]IT Department, MVSR Engineering College, Hyderabad, India
E-mail : jubairbasha@mjcollege.ac.in[1], Salman.abdul.moiz@ieee.org[2], Rizwanullah.md@gmail.com[3]

*Abstract -* One of the goals of software design is to model a system in such a way that it is easily understandable. Nowadays the tendency for software development is changing from manual coding to automatic code generation; it is becoming model-based. This is a response to the software crisis, in which the cost of hardware has decreased and conversely the cost of software development has increased sharply. The methodologies that allowed this change are model-based, thus relieving the human from detailed coding. Still there is a long way to achieve this goal, but work is being done worldwide to achieve this objective. This paper presents the drastic changes related to modeling and important challenging issues and techniques that recur in MBSD.

## I. INTRODUCTION

Model is an abstraction of some aspect of a system. Model-based software and system design is based on the end-to-end use of formal, composable and manipulable models in the product life-cycle. An emerging common thread is that modeling languages are domain-specific: they offer software developers concepts and notations that are tailored to capture essential characteristics of their application domain [1].

This paper presents the state-of-the-art of the Model-Based Software Development. Section-II presents the Model-Based Software Engineering (MBSE) and Model Centric Software Development (MCSD). The process Domain Engineering process [2] is presented with the specific domain in section –III. The purpose of DARE-COTS tool is discussed along with the scope of product lines. Section –IV highlights the research challenges in terms of Multi Aspect Modeling. Section –V introduces basics, different usages and important issues of techniques that recur in MBSD. Finally Section VI describes the related survey work concludes the paper.

## II. MODEL BASED SOFTWARE ENGINEERING

Model based Software Engineering is the idea of achieving code reuse and perform maintenance and product development through the use of software modeling technology and by splitting the production of software into two parallel engineering processes namely domain engineering and application. The system described by a model may or may not exist at the time the model is created. Models are created to serve particular purposes, for example, to present a human understandable description of some aspect of a system or to present information in a form that can be mechanically analyzed.[3,4]. Model-based development approaches can be roughly classified on the primary abstraction level of their focal software model.

Model-driven engineering (MDE) is a software development methodology which focuses on creating and exploiting domain models (that is, abstract representations of the knowledge and activities that govern a particular application domain), rather than on the computing (or algorithmic) concepts. The MDE approach is meant to increase productivity by maximizing compatibility between systems (via reuse of standardized models), simplifying the process of design (via models of recurring design patterns in the application domain), and promoting communication between individuals and teams working on the system (via a standardization of the terminology and the best practices used in the application domain).

### A. Model-Centric Software Development

The idea of using models to alleviate software complexity has been around for many years. However, researchers have largely applied models to selected elements of the development process, particularly structural and compositional aspects in the design phase and model checking and verification in the testing phase. The different stages of software development lifecycle are insufficiently interconnected with each other due to the lack of a unified way to express relevant concepts at an appropriate level of abstraction.

Model-Centric Software Development (MCSD) is an attempt at realizing a knowledge hub for the software development lifecycle. The core idea of this approach is to use models that are both concise and expressive across the development process to express the relevant concepts of each area such that they become transparent and can be used in other areas. Model-centric approaches to software development have been around for many years but it is the special field of model-driven software development dealing with the generation of executable code from implementation-level models that has stirred particular interest over the last few years. MCSD, however, encompasses a much broader scope and areas such as business process modeling, architectural models, or enterprise-wide federated repositories.

MCSD can offer a tremendous chance to leverage individual intellectual assets in software engineering in general and to fulfill Domain Driven Design's promise of business/technology alignment in particular when employed properly but can also bring a project to the brink of failure when ignoring the remarkable level of additional complexity it introduces both on the technical and organizational level.

### B. Model Driven Development and Automatic Programming

Model-driven development (MDD) typically focuses on software design models [5]. MDD and automatic programming [6] both rely on the machine to generate complete code from software artifacts of a higher-level abstraction. The difference becomes obvious if we compare source models and generated code in automatic programming and MDD. Both of them generate complete source code. However, most decisions in generated code of MDD are actually specified by designers in source models, and this is an important reason that MDD emphasizes complete and precise modeling.[7]

### III. DOMAIN ENGINEERING

Domain Engineering (DE) is a process in which the reusable component is developed and organized and in which the architecture meeting the requirements of this domain is designed. The "domain" refers to the functional areas covered by a group of application systems that have the same or similar software requirements [8].

Domain engineering process [2] is depicted in figure 1. DE consists of three main stages i.e. domain analysis, domain design and domain implementation. For Domain Analysis support, DARE-COTS tool is presented [9]. Initially, in a particular domain it is mandatory to get the universal and variable characteristics of group systems. By abstracting the characteristics, domain analysis model can be generated. Based on this model the domain specific software architecture can be designed and then reusable components will be generated and organized.

Thus, when developing a new system in new domain, we have to identify the system's requirements and specification as per the domain model, and can generate the new design as per the Domain Specific Software Architecture (DSSA), then select the particular components to assemble the new system. The process of developing an only single application system is called Application Engineering.

In [2] the method of domain engineering which is described and the DSSA of Product Quality Tracking System has been presented. The discussion shows on how to develop an open and reusable product quality tracking system on the basis of domain engineering. The research shown in this article specifies the need of reusing the major functionality of the system when the application which is developed in a similar domain for which the components are available. Further research has to focus on a specific product quality tracking system using asserts and to perfect the component repository. Massiom et al [10] evaluated an application of domain analysis in a specific domain, i.e. production management and measured the results with some improvements to be done by integrating domain analysis method in standard development.

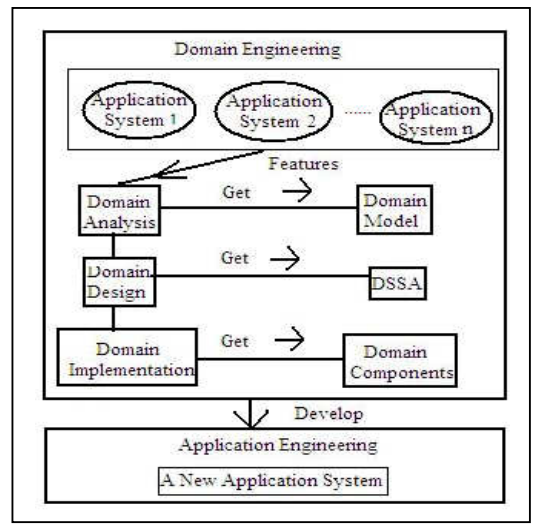

Fig.1 : Process of Domain Engineering [2]

## IV. CHALLENGES

What MBSD suggests is essentially a role transition of software models from documentation to development. This implies an enhanced requirement on software models for completeness and precision, compared with the traditional use of models. It also demands an efficient mechanism of model-implementation mapping, which is not only about generating model-prescribed code, but also about managing the consistency between model and code over the passage of time. In general, no MBSD approach can survive in the long run if the cost of model-implementation mapping significantly exceeds that of working on code directly. This section describes the research challenges in multi-aspect modeling of MBSD from the perspectives of what it is, why it is hard, and how existing mechanisms are deficient in addressing it.

### A. Multi-Aspect Modeling

Software models in the development of complex software often need to describe the system from multiple aspects, such as structure, behavior, and non-functional properties. Important research progress has been made in this area [11, 12]. However, most of existing modeling technologies are based on the assumption that software models are documentation artifacts that are peripheral to code development. With regard to structure, models such as UML class diagrams may be fine for use in MBSD. With regard to behavior, few models created with current technologies are amenable to software synthesis in MBSD; the situation with regard to non-functional models is even worse. The challenge is that software models in MBSD not only have to contain enough details to generate relatively complete code, but also need to be, and stay, simpler than the software programs created during this process.

Existing behavioral modeling methods include those that are based on formal notations and those that are more informal, but with a practical bias. None, however, provides an appropriate form for MBSD. Formal behavioral modeling methods include the use of process algebras like CSP and the pi-calculus. Providing a basis for automatic analysis is one of their main purposes. They are seldom appropriate for software development because of their limited expressiveness. In most cases, developers would rather write code directly. Examples of more informal methods include interaction diagrams, state diagrams, and activity diagrams of UML. Traditionally, these methods are mainly for communication and system comprehension. Their incompleteness properties have decided that they cannot be used alone for behavioral modeling in MDD [13], which emphasizes complete modeling. In cases where only executions of significance are concerned, such as architecture-centric development, practical methods like sequence diagrams may be a good choice after some form of extension [11].

## V. TECHNIQUES

### A. Domain Specificity

Basics : Exploiting domain specificity is primarily about developing artifacts that may be reused in developing multiple applications within a given domain. In domain-specific MBSD, reusable assets include DSLs, domain components, and reference architecture. The use of DSLs raises the level of abstraction, and improves the expressive power of software models. A library of reusable components supports software implementation through component composition. Reference architectures serve guides to the composition process. They simplify the management of supplier relationships by describing the specific contexts in which components operate.

Different usages : Domain-specific MBSD includes application generators, MIC, DSSA, and generative software development. A significant discriminator of these four approaches is the domain asset being reused. The application generator approach reuses code generators; MIC uses DSLs to model embedded systems; DSSA and generative software development both recognize reference architectures, domain components, and configuration knowledge as reusable assets. DSSA is different from generative software development because the latter uses a configuration generator to implement configuration knowledge and automate the selection of components [14], whereas this is usually done manually in DSSA. In addition, the creation of reference architecture in generative software development is primarily to identify "uses" dependencies between component categories and facilitates the implementation of components. In contrast, the DSSA approach uses reference architectures as a key element in the creation of a specialized architecture.

Issues : The exploitation of domain specificity plays a significant role in MDD, which faces the challenge of complete modeling and full code generation. What a generic MDD (e.g. MDA) does is directly specifying system (dynamic) details in software models. This not only makes models complicated and potentially degrades their usability, but also imposes a high requirement on the extensibility of the modeling language used. Domain-specific MDD [15] is much more favorable at this point. On the one hand, a DSL is more expressive than a generic modeling language (e.g. UML) when applied in a specific domain. One the other hand, reuse of domain specific code generators or components greatly reduces the amount of generated code, and thus, the information that has to be specified in software models.

*B. Metamodeling*

Basics : A Meta model is a model that is written in a meta language to define some specific modeling language [13]. In essence, meta modeling is important because it provides a means for the machine to read, write, and understand models that previously were interpreted only by people. From this perspective, met modeling plays a key role in automating MBSD. With models understandable to computers, tools can be built for model creation, code generation, and consistency management.

Different usages : Met modeling is primarily used in MDD and architecture centric software development. A representative example is MDA, which is based on OMG's four-layer meta-level hierarchy [13]. Its primary modeling language, UML, is defined by a metamodel written in MOF. Different from MDA, MIC as another MDD approach uses UML as its metalanguage to define its DSLs. In particular, MIC includes a generic modeling environment that can be customized by the metamodel of a domain language to support modeling in a given domain. At this point, it is very similar to ArchStudio [16], a metamodeling based tool for architecture-centric software development. The modeling notation used by ArchStudio is xADL, an XML-based architecture description language. Significantly, users are allowed to extend the schemas of xADL for new features. ArchStudio reads schemas and automatically generates a data-binding library for new tools.

Issues : Meta level and software abstraction level are two different concepts in MBSD. Meta level reflects the linguistic instance-of relationship between a model and its metamodel. In other words, a model is written in a language that is defined by the models metamodel at a higher meta level. In contrast, software abstraction level characterizes a software model in terms of to what extent it hides unimportant information to a software developer. For example, the abstraction provided by software architecture allows a software architect to focus on principal design decisions without worrying about implementation details. From this perspective, meta level and abstraction level are orthogonal concepts.

*C. Iterative Transformation*

Basics : Iterative transformation is extensively used in transformational programming. The central idea is to break a transformation that crosses an abstraction gap into sufficiently small steps, so that each step generates another representation that is easier to implement than the first. What this means in the context of MBSD is an incremental way to map source models into implementations, especially when source models are too abstract to directly generate code from.

Different usages : Style-based architecture refinement is just a typical application of this idea. It maps an abstract architecture into a concrete architecture through a series of small transformations, each of which involves the application of a preproved transformation pattern that is specific to an architecture style. Software Factories use a similar approach, so called progressive transformation, to map domain-specific models into implementations. Layers of simplifying abstractions are successively generated during this process. Another less obvious example is MDA, where the use of PSM to facilitate the mapping of PIM to a working implementation on a middleware platform actually reflects the same spirit of iterative transformation.

Issues : The applications of iterative transformation presented above are all limited to certain ranges, such as a specific architecture style, an application domain, or a middleware platform. In addition, their source and generated models usually stay close in terms of conceptual level. At this point, we think this represents proper uses of iterative transformation. Not only the development portion that can be pre-planned and specified is increased, but also the complexity level is reduced. This is different from automatic programming discussed in Section II, which assumes software development can be pre-planned in a generic way and, in general, faces a significant conceptual gap between requirements specifications and executable programs.

## VI. CONCLUSION

The literature of [6,7,17] specifically discuss the advantages, disadvantages, difficulties and facilities of MDD, an important branch of MBSD. This paper discusses the techniques for the challenging issues in the MBSD. Especially, Section V throws so many opportunities for the research issues in future directions. As a part of our research work we will consider this MBSD as an approach for the development of Domain Specific Components that will leads to the Generic MDD. The issues and challenges presented in this paper are useful for the research initiators to carry out further research in MBSD. By considering these issues we can design a model which is reusable. A Generic Framework has to be developed.

## ACKNOWLEDGMENT

The work was partly supported by the R & D Cell of Muffakham Jah College of Engineering & Technology, Hyderabad, India. The authors would like to thank to all the people from Industry and Academia for their active support.